\documentclass[
 reprint, 10pt,
 amsmath,amssymb,
 aps, superscriptaddress
]{revtex4-1}

\usepackage{graphicx}
\usepackage{dcolumn}
\usepackage{hyperref}
\usepackage[dvipsnames]{xcolor}
\usepackage{bm}
\usepackage{bbold}
\usepackage{bbm}

\definecolor{marie}{RGB}{0,128,128}
\definecolor{todo}{RGB}{102,51,153}
\definecolor{emph}{RGB}{0,150,150}
\begin{document}

\title{Strong coupling in molecular systems: a simple predictor employing routine optical measurements}

\author{Marie S Rider}
\email{M.S.Rider@exeter.ac.uk}
\affiliation{Department of Physics and Astronomy, Stocker Road, University of Exeter, Devon EX4 4QL, United Kingdom}
\author{Edwin C Johnson}
\affiliation{Department of Chemistry, University of Sheffield, Brook Hill, Sheffield, S3 7HF, United Kingdom}
\author{Demetris Bates}
\affiliation{Department of Chemistry, University of Sheffield, Brook Hill, Sheffield, S3 7HF, United Kingdom}
\author{William P Wardley}
\affiliation{Department of Physics and Astronomy, Stocker Road, University of Exeter, Devon EX4 4QL, United Kingdom}
\author{Robert H Gordon}
\affiliation{Department of Physics and Astronomy, University of Sheffield, Hicks Building, Hounsfield Road, Sheffield, S3 7RH, United Kingdom}
\author{Robert D J Oliver}
\affiliation{Department of Physics and Astronomy, University of Sheffield, Hicks Building, Hounsfield Road, Sheffield, S3 7RH, United Kingdom}
\affiliation{Department of Materials Science and Engineering, University of Sheffield, Sir Robert Hadfield Building, Mappin Street, Sheffield, S1 3JD, United Kingdom}
\author{Steven P Armes}
\affiliation{Department of Chemistry, University of Sheffield, Brook Hill, Sheffield, S3 7HF, United Kingdom}
\author{Graham J Leggett}
\affiliation{Department of Chemistry, University of Sheffield, Brook Hill, Sheffield, S3 7HF, United Kingdom}
\author{William L Barnes}
\email{W.L.Barnes@exeter.ac.uk}
\affiliation{Department of Physics and Astronomy, Stocker Road, University of Exeter, Devon EX4 4QL, United Kingdom}

\date{\today}
\begin{abstract}
We provide a simple method that enables readily acquired experimental data to be used to predict whether or not a candidate molecular material may exhibit strong coupling. Specifically, we explore the relationship between the hybrid molecular/photonic (polaritonic) states and the bulk optical response of the molecular material. For a given material this approach enables a prediction of the maximum extent of strong coupling (vacuum Rabi splitting), irrespective of the nature of the confined light field. We provide formulae for the upper limit of the splitting in terms of the molar absorption coefficient, the attenuation coefficient, the extinction coefficient (imaginary part of the refractive index) and the absorbance. To illustrate this approach we provide a number of examples, we also discuss some of the limitations of our approach.
\end{abstract}


\maketitle

\section{Introduction}
\begin{figure*}
\includegraphics[width=\textwidth]{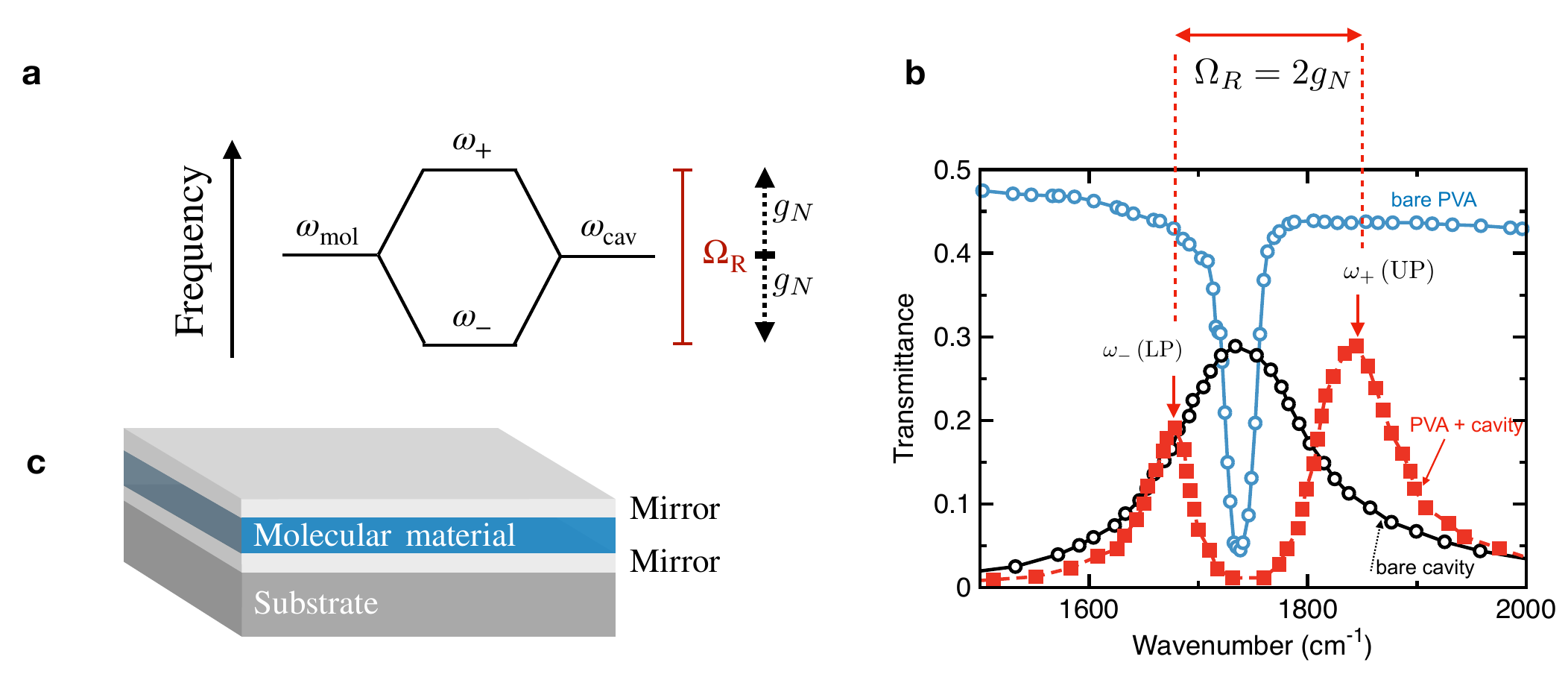}
\caption{(a) Strong coupling between a molecular resonance and a cavity resonance results in two new hybrid energy levels separated by frequency $\Omega_\mathrm{R} = \omega_+ - \omega_- = 2g_N$. (b) This hybridization and resultant energy level splitting can be seen in the transmission of the system, here employing data adapted from Shalabney \textit{et al.}~\cite{Shalabney_NatComm_2015_6_5981}. The blue data are the transmittance of a polymer (polyvinyl acetate, PVA) film in the absence of a cavity. The strong absorption at $\sim$ 1730 cm$^{-1}$ is due to the C=O stretch mode in the PVA. The black data show the transmittance of an empty cavity, i.e. one in which the oscillator strength associated with the molecular resonance has been set to zero. The red data show that when the cavity is filled with PVA the original single transmission peak is lost and two new transmittance peaks emerge, the upper ($\omega_+$) and lower ($\omega_-$) polaritons. (c) The molecule-cavity system we consider is that of a polymer sandwiched between two mirrors.}
\label{fig:introduction}
\end{figure*}

Coupling of a vibrational or an excitonic resonance of an ensemble of molecules to an optical cavity mode (confined light field) has become a major area of research. Interest is focused in particular on what happens when the coupling between the molecules and the confined light field enters the strong coupling regime~\cite{Ebbesen_ACS_Accounts_2016_49_2403,Rider_CP_2021_62_217}. In this regime the molecular resonance at (angular) frequency $\omega_0$ hybridises with an optical resonance (e.g. that of an optical microcavity~\cite{Lidzey_PRL_1999_82_3316}) to produce two new modes -- polaritons -- that are part light, and part molecular resonance, at frequencies $\omega_+$ (upper polariton) and $\omega_-$ (lower polariton). The difference in  frequency (energy) of the upper and lower polaritons at resonance is usually called the Rabi-frequency and, since we often wish to know how far from the original molecular resonance, $\omega_0$, the two polaritons $\omega_{\pm}$ are, we write $\Omega_R=\omega_+-\omega_-=2g_N$ where $g_N$ is the N-molecule interaction strength~\cite{Rider_CP_2021_62_217}, see FIG.~\ref{fig:introduction}.
The extent of the hybridisation can be so dramatic that the associated energy levels may be shifted by a substantial fraction of the unperturbed resonance energy~\cite{Kockum_NatRevPhys_2019_1_19}. Interest is particularly strong from the perspective of using molecular strong coupling to modify energy transport between molecules~\cite{Zhong_AngChemIntEd_2016_55_6202,Georgiou_ACSPhot_2018_5_258}, to control exciton transport~\cite{Balasubrahmaniyam_NatMat_2023_22_338}, and to alter chemical reactions~\cite{Ebbesen_ACS_Accounts_2016_49_2403}, although the extent to which this is possible is still an area of vigorous debate~\cite{sidler2020polaritonic,hirai2020recent,li2021cavity,campos2022swinging,chen_2301.19133,Khazanov_ChemSocRev_2023_4_041305}.\\

Various theoretical and computational approaches can be adopted both to explore and understand the associated phenomena~\cite{herrera2020molecular,fregoni2022theoretical}, ranging from simple classical oscillator models~\cite{Rider_CP_2021_62_217} to sophisticated macroscopic quantum electrodynamics~\cite{feist2021macroscopic} and ab-initio numerical quantum chemistry approaches~\cite{sokolovskii2022enhanced,berghuis2022controlling}. Classical models are remarkably effective, and simple mechanical coupled oscillator models and transfer matrix approaches are particularly prevalent in the literature, and often seem to successfully account for observed phenomena~\cite{zhu1990vacuum,weisbuch1992observation}.\\

Both the dispersion of the split modes (polaritons) and the extent of the Rabi splitting (shown in FIG.~\ref{fig:introduction}b) can be predicted solely on the basis of a detailed knowledge of the bulk optical response of the molecular material involved, for example that of the dye-doped polymer used to fill the cavity shown in FIG.~\ref{fig:introduction}(c). This approach to modelling strong coupling has been successfully used in the context of phonon resonances~\cite{Barra-Burillo_NatComm_2021_12_6206}. Here we investigate these ideas in the context of molecular excitonic and vibrational resonances, for example dye-doped polymers. We focus our attention on how the results of simple optical measurements can be harnessed in a predictive way to help in the design of molecular strong coupling systems. To keep the message of the main text clear, we restrict the majority of derivations to the appendices. 

\section{Summary of key formulae}

\subsection{Strong coupling criterion}

Whether or not one can expect strong coupling to be observed depends on whether the interaction strength (rate) between the molecular resonators and the confined light field exceeds the dissipation rates. We can also consider this condition from a spectroscopic viewpoint: we need the Rabi frequency, $\Omega_{\textrm{R}}$, to exceed the mean of the cavity, $K$, and molecular, $\Gamma$, linewidths. Noting that the Rabi frequency is twice the interaction strength, then for strong coupling we have the following condition~\cite{Rider_CP_2021_62_217},
\begin{align}
\Omega_R=2g_N > \frac{\Gamma+K}{2}.
\label{eq:spectral sc condition}
\end{align}
A fuller discussion of strong coupling criteria is given in appendix \ref{app:criteria}.\\

We focus here on calculating a value for $g_N$ from experimentally measurable parameters. We can see from \eqref{eq:spectral sc condition} that strong coupling is possible provided $\Omega_R-\Gamma$ is positive; if it is then we can rearrange equation \eqref{eq:spectral sc condition} to set a \textit{usable} upper bound on the cavity linewidth, $K$, for which strong coupling can be observed, specifically,
\begin{align}
K < 2\Omega_R - \Gamma.
\label{eq:cavity linewidth limit}
\end{align}
%

\subsection{Interaction strength}

The strength of the interaction of $N$ molecular resonators and a cavity mode, $g_N$, is based on the electric dipole interaction.  We consider molecular resonances that involve electric dipole transitions, at angular frequency, $\omega_0$, and that have a transition dipole moment, $\mu$. Note that the formation of polariton modes involves the interaction of this dipole moment with the cavity vacuum field, $E_{\textrm{vac}}$. No external source of light is involved in the formation of these new, coupled modes. Observation of the modes will usually involve light, but the modes exist whether or not the cavity is illuminated~\cite{Ebbesen_ACS_Accounts_2016_49_2403,Rider_CP_2021_62_217}.\\

The interaction energy for a single electric dipole is given by,
\begin{align}
\hbar g = \mu.E_{\textrm{vac}},
\label{eq:single dipole}
\end{align}
where we have assumed that the dipole moment and field are aligned. The (RMS) strength of the vacuum field is $\sqrt{\hbar\omega_0/2V_m\varepsilon_0\varepsilon_{\textrm{host}}}$, where $\varepsilon_{\textrm{host}}$ is the background permittivity of the molecular material, and $V_m$ is the volume of the cavity mode~\cite{Rider_CP_2021_62_217}. For example, the mode volume associated with the plasmon mode of a gold nanosphere will be roughly the surface area of the sphere multiplied by the decay length of the electric field associated with the plasmon mode into the surrounding medium~\cite{Barnes_AmJPhys_2016_84_593}. As another example, the mode volume associated with an optical microcavity will be roughly the cavity thickness multiplied by the area of the mode, which in turn is dictated by the spatial coherence properties of the mode~\cite{Ujihara_JapJAppPhys_1991_30_L901}. Calculating exact mode volumes is a subtle business~\cite{Lalanne_LPR_2018_12_1700113}, but we only need to appreciate the underlying ideas involved here, there is also the question of whether the molecules of interest fill the mode volume, a matter we discuss in the third bullet point bellow. The last piece of information we need is that the interaction energy scales as the square root of the number of dipoles (molecules) involved~\cite{Rider_CP_2021_62_217}. We thus have,
\begin{align}
g_N = \left(\frac{\Omega_R}{2}\right) = \frac{1}{\hbar}\sqrt{N} \mu\,E_{\textrm{vac}} = \frac{1}{\hbar}\sqrt{N} \mu\,
\sqrt{\frac{\hbar \omega_0}{2V_m\,\varepsilon_0\,\varepsilon_{\textrm{host}}}}.
\label{eq:interaction_1}
\end{align}
Whether or not this value of $g_N$ is sufficient for strong coupling to be observed depends on whether the strong coupling condition, equation \eqref{eq:spectral sc condition} is met.\\

Equations \eqref{eq:spectral sc condition} and \eqref{eq:interaction_1} are standard results, but they are not convenient if one wishes to make an estimate as to whether a particular molecular resonance/material combination will yield strong coupling. To see why we can rearrange equation~\eqref{eq:interaction_1} to write the coupling energy, $E_C$, in terms of the coupling strength $g_N$ as,
\begin{align}
E_C = \hbar g_N = \sqrt\frac{\mu^2 E_0 N}{2 \varepsilon_0 \varepsilon_{\textrm{host}} V_m},
\label{eq:interaction_2}
\end{align}
where $E_0$ is the energy of the molecular resonance, and is assumed to be matched to the resonance frequency of the cavity. Our result, \eqref{eq:interaction_2}, is similar to that of~\cite{Agranovich_PRB_2003_67_085311,Tsargorodska_NL_2016_16_6850}; we note however that the exact form of this expression relies on the system of units used, the assumed spatial variation of the cavity field, and the distribution and orientation of the molecules within the cavity.
If we know the dipole moment, $\mu$, and we also know the number density of molecular resonators in the cavity, $N/V_m$, then the interaction strength is easily determined. However $N/V_m$, and in particular $\mu$, are not easily derived from standard experimental measurements, although it can be done, see Appendix~\ref{app:dipole} and also~\cite{Barnes_JOpt_2024}.\\

In what follows we reformulate equation \eqref{eq:interaction_1} in a number of ways so as to more easily facilitate strong coupling predictions based on common measurements of material properties.\\

\renewcommand{\arraystretch}{0.5}
\begin{table*}[htb!]
\small
\centering
\resizebox{\textwidth}{!}{%
\begin{tabular}{| l || l | l | l |}
\hline

 & $\bar\nu$ (cm$^{-1}$)& $\omega$ (rad s$^{-1}$)& $E$ (eV)\\
\hline
\hline
&&&\\
$\epsilon$ &$\cfrac{0.24}{\sqrt{n_{\textrm{host}}}}\sqrt{\epsilon_{max}C_m\delta\bar{\nu}}$ & $\cfrac{1.05\times 10^{5}}{\sqrt{n_{\textrm{host}}}}\sqrt{\epsilon_{max}C_m\delta\omega}$ & $\cfrac{2.70\times 10^{-3}}{\sqrt{n_{\textrm{host}}}}\sqrt{\epsilon_{max}C_m\delta E}$  \\
&&&\\
\hline
\hline
&&&\\
$\alpha$ &$\cfrac{0.16}{\sqrt{n_{\textrm{host}}}}\sqrt{\alpha_{max}\delta\bar{\nu}}$ & $\cfrac{6.92\times 10^{4}}{\sqrt{n_{\textrm{host}}}}\sqrt{\alpha_{max}\delta\omega}$ & $\cfrac{1.78\times 10^{-3}}{\sqrt{n_{\textrm{host}}}}\sqrt{\alpha_{max}\delta E}$  \\
&&&\\
\hline
\hline
&&&\\
$\kappa$ &$\cfrac{0.56}{\sqrt{n_{\textrm{host}}}}\sqrt{\kappa_{max}\,\nu_0\,\delta\bar{\nu}}$ & $\cfrac{0.56}{\sqrt{n_{\textrm{host}}}}\sqrt{\kappa_{max}\omega_0\,\delta\omega}$ & $\cfrac{0.56}{\sqrt{n_{\textrm{host}}}}\sqrt{\kappa_{max}E\,\delta E}$  \\
&&&\\
\hline
\hline
&&&\\
$\textit{a}$ &$\cfrac{0.24} {\sqrt{n_{\textrm{host}}}}\sqrt{\cfrac{\textit{a}_{max}\delta\bar{\nu}}{l}}$ & $\cfrac{1.05\times 10^{5}}{\sqrt{n_{\textrm{host}}}}\sqrt{\cfrac{\textit{a}_{max}\delta\omega}{l}}$ & $\cfrac{2.70\times 10^{-3}}{\sqrt{n_{\textrm{host}}}}\sqrt{\cfrac{\textit{a}_{max}\delta E}{l}}$  \\
&&&\\
\hline

\end{tabular}}
\caption{\textbf{Formulae for the maximum interaction strength, $g_N$, in terms of material parameters}. Equations are given (by row) for four different material parameters: molar absorption coefficient $\epsilon$; absorption coefficient $\alpha$; extinction coefficient $\kappa$; and absorbance (\textit{a}) (equivalent to optical density (OD)), and by column for different units: wavenumber $\bar\nu$ (cm$^{-1}$); angular frequency $\omega$ (rad s$^{-1}$); and energy $E$ (eV). In each column the prefactors for the equations are set so as to yield the coupling strength in the same units. Thus, for the wavenumber column, using the equations given above will yield a coupling strength in wavenumbers (cm$^{-1}$) etc.. It is important to note the units used here, we have tried to adopt the conventions used in practice. Accordingly, for the top row the molar absorption coefficient, $\epsilon$, is in units of dm$^{3}$ mol$^{-1}$ cm$^{-1}$, whilst the concentration, $C$, is in units of mol dm$^{-3}$. For the second row the attenuation coefficient is in units of cm$^{-1}$. For the third row the extinction coefficient, $\kappa$, is dimensionless and has no units. In the fourth row the absorbance, $a$, is also dimensionless and so has no units. In this row the sample path length $l$ has units of cm. Note also the assumptions concerning these equations, listed in the main text (dipole orientation, line broadening and mode volume).}
\label{tab:params}
\end{table*}

We show in Appendices~\ref{app:mat} and~\ref{app:ext}, that the interaction strength in terms of the extinction coefficient $\kappa$ (equal to the imaginary part of the complex refractive index) of a thin solid layer of material (or a cuvette of material in solution) can be found using the following procedure:
\begin{itemize}
    \item We assume that the molecular resonance can be described as a Lorentzian oscillator (LO).
    \item We use the LO model to write the interaction strength in terms of an oscillator strength for the transition, rather than a dipole moment.
    \item The LO model then allows us to obtain the permittivity of the material, $\varepsilon(\omega)$, and from $\varepsilon(\omega)$ we can extract the extinction coefficient (the imaginary part of the (complex) refractive index $n+i\kappa$) through $(n+i\kappa)^2=\varepsilon$.
\end{itemize}
Details are given in Appendices~\ref{app:mat} and~\ref{app:ext}; the result is that the interaction strength in terms of the peak (maximum) value of the extinction coefficient $\kappa$ can be approximated as,
\begin{align}
\label{eq:gn extinction}
g_N=\frac{0.56}{\sqrt{n_{\textrm{host}}}}\sqrt{\kappa_{\textrm{max}}\,\omega_0\,\delta\omega},
\end{align}
where, for clarity, $\kappa_{\textrm{max}}$ is the maximum value of the extinction coefficient associated with the molecular transition, $\omega_0$ is the (angular) frequency of the molecular transition (rad $s^{-1}$), $\delta\omega$ is the width of the extinction feature (rad $s^{-1}$), and $n_{\textrm{host}}$ is the background refractive index of the molecular host (e.g. solvent or polymer).
For practical purposes -- see Appendix~\ref{app:criteria} -- we take $\Gamma\approx\delta\omega$ if we are working in rad $s^{-1}$, and $\Gamma\approx\delta \bar{\nu}$ or $\Gamma\approx\delta E$ if we are working in cm$^{-1}$ or eV respectively.\\

We note a number of assumptions that we have made:
\begin{itemize}
    \item We have assumed that the dipole moments of the molecules are randomly oriented. If instead they were oriented, for example, in the direction of the E-field of the light, then we would need to multiply the right-hand side of \eqref{eq:gn extinction} by $\sqrt{3}$.
    \item The question of how best to incorporate inhomogeneous broadening is on-going, e.g. via various modified forms of the LO model~\cite{on2017effect,djurivsic2000modeling,dutta2022modeling}. In appendix \ref{app:ext} the final form of equation \eqref{eq:gn extinction} is based on relaxing the assumption that the description of the transition should be Lorentzian, we use a generalised spectroscopic approach~\cite{Turro}.
    \item We have also implicitly assumed that the molecules of interest fill the mode volume. For a closed Fabry-P{\`e}rot cavity this may be a good approximation, but in other situations this may not be the case, e.g. a monolayer of molecules on a metal surface that supports a surface plasmon mode, or on a metallic particle~\cite{Zhengin_JPCC_2016_120_20588}. Investigating the effect of only partially filling the mode volume is beyond the scope of the present study; our calculations provide a `best case' in this regard.
\end{itemize}
We note that Gunasekaran \textit{et al.} have used a related approach to extract predictive information from absorption spectra~\cite{Gunasekaran_2308.08744}. In Table~\ref{tab:params} we provide equivalent formulae in terms of a number of different common spectral parameters, and we do so for a variety of spectral units: (i) wavenumber $(\bar\nu)$, in cm$^{-1}$; (ii) angular frequency $\omega$, in radians s$^{-1}$, and (iii) electron volts (eV); details are given in the appendices. Note that these parameters are all frequencies, or equivalent. Frequently experimental data are acquired in terms of wavelength, e.g. from a UV-VIS (ultraviolet-visible) spectrophotometer. In this case the data need to be converted to frequency to avoid lineshape distortion~\cite{Mooney_JPCL_2013_4_3316}; furthermore, making direct use of transmittance to determine the width is likely to lead to errors. The three assumptions listed immediately above also apply to the formulae listed in table ~\ref{tab:params}. In the next section we look at a number of worked examples.

\section{Worked Examples}
\label{sec:worked_examples}

\subsection{Absorbance: Nile Red in a polymer brush film}

We consider first a system consisting of the dye Nile Red attached to a surface-grafted aldehyde-functional hydrophilic polymer brush scaffold. 
Brushes are thin films (typically $<$ 100 nm in thickness) in which polymer chains are end-tethered to an underlying substrate~\cite{Brittain2007}. These surface layers have been shown to have many desirable properties~\cite{Stuart2010}, including increased antifouling~\cite{Stuart2010} and lubricity~\cite{Klein1991} and can act as a scaffold for various small and large molecules~\cite{Hui2013,Gautrot_brushreview2023}.
In the example presented here, a hydrophilic aldehyde functional polymer brush (PAGEO5MA) was grown from a glass cover slip~\cite{Brotherton2021,Brotherton2023,Johnson2023}, with a dry thickness = 40 nm, before decoration with an amino-functionalised Nile Red analogue (2-(2-Aminoethoxy)Nile Red)~\cite{Tosi2016}.
We have shown that PAGEO5MA can be conjugated with reactive amines through reductive amination chemistry to produce highly functionalised coatings ($>$ 80\% of available reactive sites)~\cite{Brotherton2021,Brotherton2023,Johnson2023}.
Further details of the synthesis and characterisation of the Nile Red analogue, the PAGEO5MA brush and Nile Red-brush system are outlined in Appendix~\ref{app:Methods}.\\

\begin{figure}
\includegraphics[width=1.0\columnwidth]{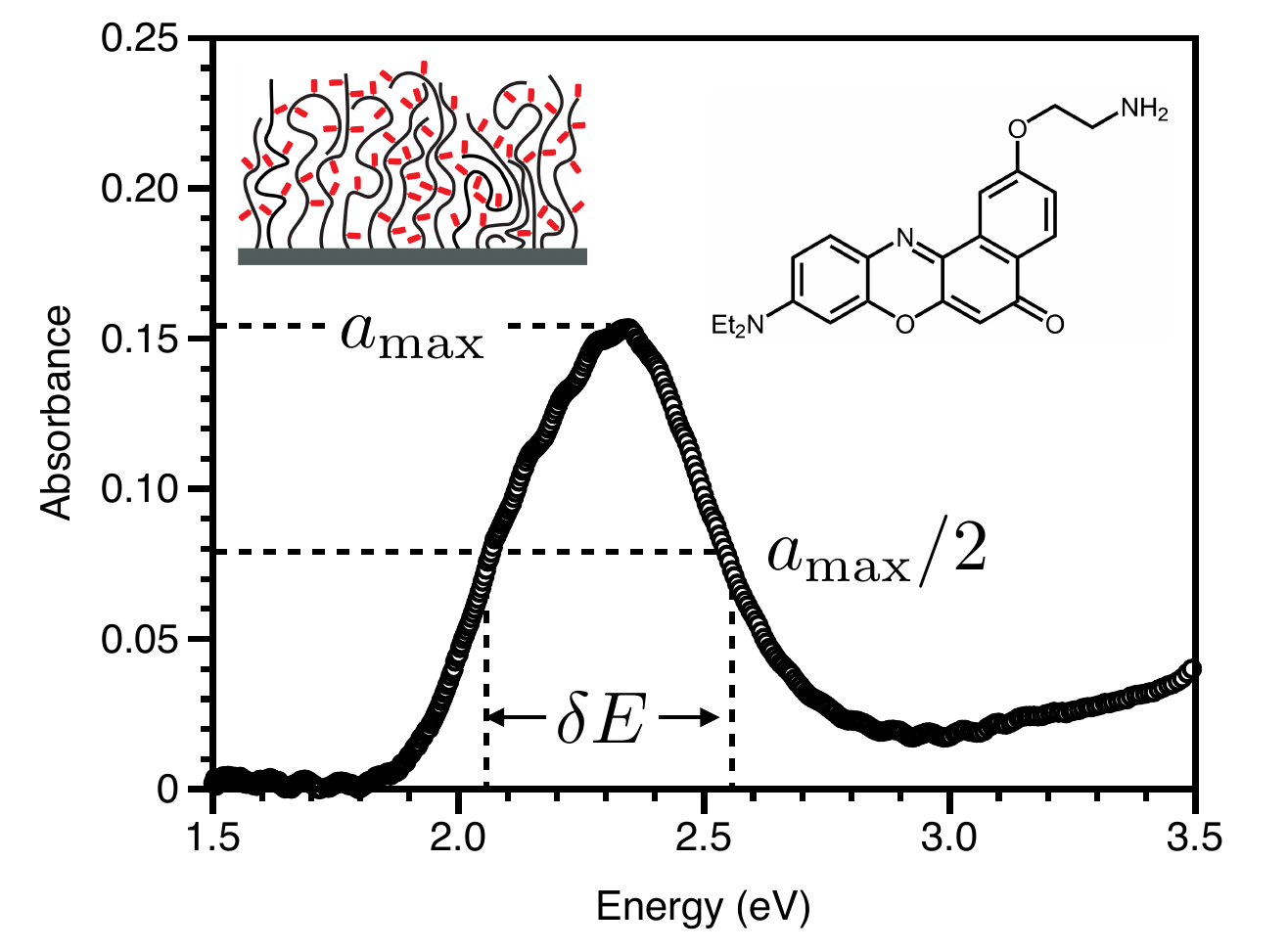}
\caption{\textbf{Absorbance of Nile Red functionalised PAGEO5MA polymer brush} Experimental data for the absorbance of a 40 nm PAGEO5MA polymer brush functionalised with Nile Red. From these data we find the maximum absorbance to be $a_{\textrm{max}}\sim$ 0.15, and the width to be $\delta E \sim$ 0.49 eV. The inset top left shows a schematic of the dye molecules (red) incorporated into the brush. The inset top right shows the structure of the variant of Nile Red we used (for further details see Appendix \ref{app:Methods}).}
\label{fig:Nile Red}
\end{figure}

The transmittance is measured in a standard UV-VIS spectrophotometer. Usually the absorbance is calculated from the transmittance as $\textit{a} = \textrm{log}_{10}\,({\textrm{I}(0)}/{\textrm{I}(t)})$, where $\textrm{I}(0)$ and $\textrm{I}(t)$ are the incident and transmitted intensities respectively. However, this equation does not take into account and reflection of light by the sample, and for a highly absorbing thin film, this reflection can be substantial (in the case of the film used for figure \ref{fig:Nile Red} the maximum reflection - on resonance - was $\sim$ 10\%) leading to an overestimate of the absorbed power. To account for this reflected light we need to use,

\begin{equation}
\label{eq:abs}
\textit{a} = \textrm{log}_{10}\,\left({\frac{\textrm{I}(0)-\textrm{I}(r)}{\textrm{I}(t)}}\right).
\end{equation}

\noindent where $\textrm{I}(r)$ is the reflected intensity, see figure \ref{fig:Beer-Lambert} below.

\noindent The measured absorbance of such a thin film on a glass substrate is shown in FIG.~\ref{fig:Nile Red}. Note that for these data the power lost to reflections was estimated by simulating the reflectance using a Fresnel-type calculation, a calculation that incorporated the resonance of the Nile Red dye. If this correction is not made the absorbance is overestimated by 15\%. Further note that we have also ignored any possible scattering. In a future publication will discuss these important issues concerning data analysis from UV-VIS spectrophotometer measurements more fully.
Since the data are in terms of eV, see FIG.~\ref{fig:Nile Red}, the formula we need for the coupling strength is, see Table~\ref{tab:params}, lower right entry,
\begin{equation}
g_N=\frac{2.7\times 10^{-3}}{\sqrt{n_{\textrm{host}}}}\sqrt{\frac{\textit{a}_{max}\delta E}{l}}.
\end{equation}
From FIG.~\ref{fig:Nile Red} the maximum absorbance is 0.15, and the width is $\delta E \sim$ 0.49 eV. The polymer brush is $l$ = 40~nm thick but the units we need here are cm, so the thickness is $4\times 10^{-6}$ cm, thus with $\sqrt{n_{\textrm{host}}}\,\sim 1.2$, we find $g_N=$ 0.31 eV. Using this result, and the fact that $\Gamma\,\sim\,\delta E\,\sim$ 0.49 eV, together with condition \eqref{eq:cavity linewidth limit} we can see that provided  $K<0.75$ eV, i.e. we employ a cavity mode of width no greater than 0.75 eV, then observing the effects of strong coupling, e.g. split peaks (see FIG.~\ref{fig:introduction}) should be possible. This estimate assumes that the mode volume is filled by the dye, i.e. that the 40 nm thick brush layer fills the mode volume. As noted above, this may be reasonable e.g. for a particle plasmon resonance~\cite{Fofang_NL_2008_8_3481,Gentile_JOpt_2017_19_035003} but will not be the case for a surface plasmon on a planar surface, or a standard Fabry-P{\`e}rot cavity mode. Nonetheless, in these latter cases there is considerable margin to make use of the Nile Red doped brush since it would be easy to employ a cavity mode with a much narrower mode, for example a Fabry-P{\`e}rot cavity might typically have $K\,\sim$ 0.1 eV.

\subsection{Extinction: TDBC in a layer-by-layer film}

Our second example is based on a molecular system that has been a workhorse in strong coupling experiments, the aggregated dye TDBC~\cite{Bellessa_PRL_2004_93_036404,Balasubrahmaniyam_NatMat_2023_22_338}. This dye has been extensively used because it has a strong (high oscillator strength) yet narrow transition. Various approaches to making structures containing TDBC can be employed, here we show data based on ellipsometry of 4 layers of TDBC deposited on a glass substrate using a layer-by-layer approach~\cite{Bradley_AdvMat_2005_17_1881,Bradley_PRB_2010_82_033305,Balasubrahmaniyam_NatMat_2023_22_338}.\\

\begin{figure}
\includegraphics[width=1.0\columnwidth]{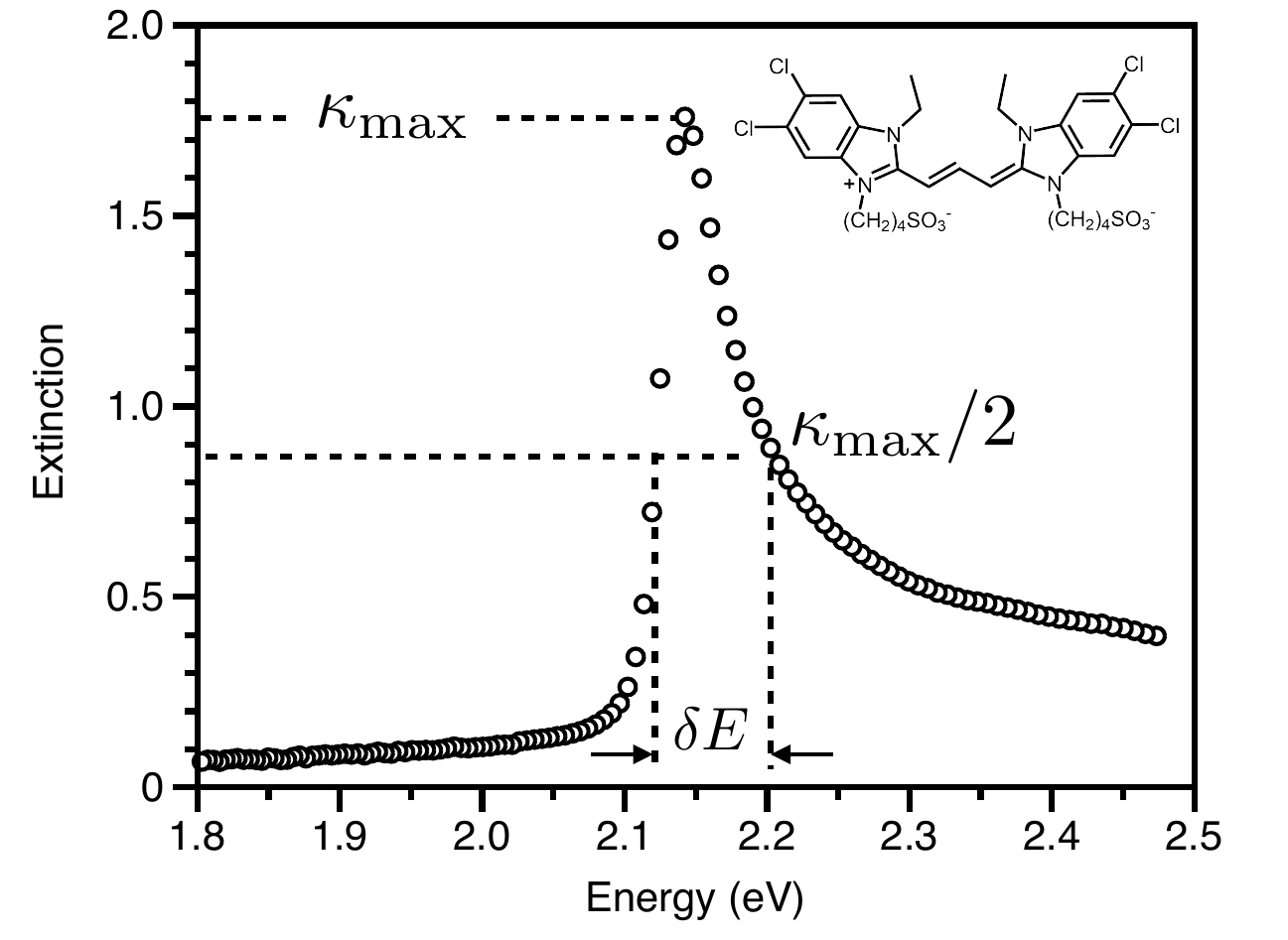}
\caption{\textbf{Real and imaginary (extinction) components of the refractive index of a TDBC} layer-by-layer sample, 4 layers thick. The data are derived from ellipsometry measurements. From these data we can estimate the transition energy to be $E$ = 2.14 eV, the peak extinction to be $\kappa_{\textrm{max}}$ = 1.75, and the FWHM of the extinction to be $\delta E$ = 0.08 eV. The inset shows the chemical structure of the TDBC monomer.}
\label{fig:TDBC}
\end{figure}

Again, because these data are in terms of eV, see FIG.~\ref{fig:TDBC}, the formula we need for the coupling strength is in row three, right hand column of Table~\ref{tab:params},
\begin{equation}
g_N = \frac{0.56}{\sqrt{n_{\textrm{host}}}}\sqrt{\kappa_{max}E\,\delta E}.
\end{equation}
From FIG.~\ref{fig:TDBC} the maximum extinction is $\kappa\,\sim$ 1.75, the width is $\delta E \sim$ 0.08 eV, and the transition is at $E \sim$ 2.14 eV. Again assuming $\sqrt{n_{\textrm{host}}}\,\sim 1.2$, we find $g_N=$ 0.25 eV. This is comfortably greater than the TDBC linewidth $\Gamma\,\sim\,\delta E \sim$ 0.08 eV so that strong coupling should be possible in a range of cavity structures. The restriction we find for the cavity linewidth using equation \eqref{eq:cavity linewidth limit} is $K<$ 0.42 eV. Note that: (i) the extinction feature is not symmetric, due to a transition on the high energy side of the main transition~\cite{Gentile_JOpt_2017_19_035003}, our estimate of the width is thus an overestimate, we might expect something closer to 0.06 eV; (ii) a significantly greater density of TDBC aggregates can be achieved than has been accomplished here, see for example~\cite{Gentile_JOpt_2016_18_015001}.

\subsection{Molar absorption coefficient: the C=O vibrational stretch transition}

\begin{figure}
\includegraphics[width=1.0\columnwidth]{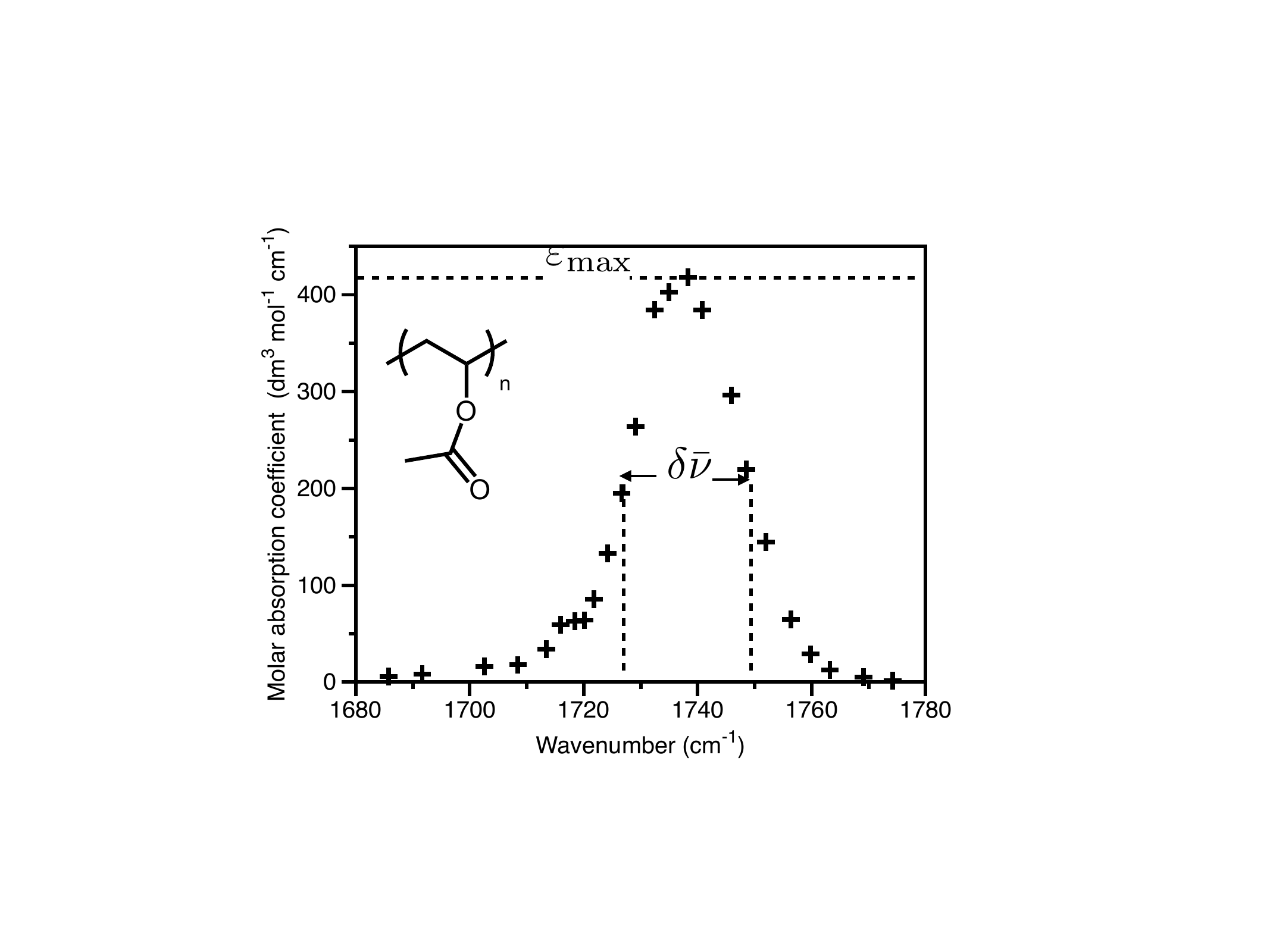}
\caption{\textbf{Experimental data for the molar absorption coefficient of a thin film of PVA on a Ge substrate}. The data are adapted from~\cite{Shalabney_NatComm_2015_6_5981}, and have been corrected for substrate reflections etc. See also~\cite{Barnes_JOpt_2024}. The inset shows the chemical structure of the PVA repeat unit.}
\label{fig:C=O transmittance}
\end{figure}

Our third example involves the C=O vibrational stretch transition in the polymer PVA. We make use of literature data from Shalbney \textit{et al.}~\cite{Shalabney_NatComm_2015_6_5981} (who measured the infrared transmittance associated with a thin (1.7 $\mu$m) spun film of PVA on a Ge substrate) to plot the molar absorption coefficient for this thin film, shown in FIG.~\ref{fig:C=O transmittance}.
As noted in Appendix~\ref{app:Beer-Lambert}, the molar absorption coefficient~\footnote{Note that this is often called the molar extinction coefficient, but molar absorption coefficient is the correct term, see~\cite{Braslavsky_PAC_2007_79_293}}, $\epsilon$ is easily related to the transmittance, and is given by,
\begin{equation}
\label{eq:mol_abs_coeff}
\epsilon = \rm{log}_{10}\left(\frac{I_0-I_{\rm{r}}}{I_{\rm{t}}}\right)\frac{1}{\textit{l}\,C_m},
\end{equation}
where again, $I_0$, $I_{\rm{r}}$ and $I_{\rm{t}}$ are the incident, reflected and transmitted intensities. As with the Nile Red data in the first worked example, account has been taken of the reflected power. By convention the units for the path length, $l$, are cm, whilst the units for the molecular concentration, $C_m$, are moles per dm$^{-3}$, i.e. moles per litre; the units for $\varepsilon$ are thus dm$^{3}$ mol$^{-1}$ cm$^{-1}$.\\

From the experiment of Shalbney \textit{et al.}~\cite{Shalabney_NatComm_2015_6_5981} we know the length $l$ in this case to be $l=1.7\times 10^{-4}$ cm (the thickness of the PVA film), whilst the concentration of C=O bonds in PVA can be evaluated using available data~\cite{polymerdatabase} as 13.8 mol dm$^{-3}$, see also~\cite{Barnes_JOpt_2024}. Using this information we can convert the transmittance data of Shalbney \textit{et al.} into a molar absorption  coefficient, see FIG.~\ref{fig:C=O transmittance}. We find $\epsilon_{max}\,\sim$ 426 dm$^{3}$ mol$^{-1}$ cm$^{-1}$. Since the data are in terms of wavenumber (cm$^{-1}$), see FIG.~\ref{fig:C=O transmittance}, the formula we need for the coupling strength is, see Table~\ref{tab:params}, top left entry,
\begin{equation}
g_N = \frac{0.24}{\sqrt{n_{\textrm{host}}}}\sqrt{\epsilon_{max}C_m\delta\bar{\nu}}
\end{equation}
From FIG.~\ref{fig:C=O transmittance} the width is $\delta \bar\nu \sim$ 21 cm$^{-1}$ so that in this case, with $\sqrt{n_{\textrm{host}}}\,\sim 1.2$, we find, $g_N=$ 70 cm$^{-1}$ so that $\Omega_R=$ 140 cm$^{-1}$; similar to the experimentally measured value of $\Omega_R=170~$cm$^{-1}$. With $\Gamma\,\sim\delta{\bar{\nu}}\,\sim$ 21 cm$^{-1}$, equation \eqref{eq:cavity linewidth limit} puts an upper linewidth on the cavity mode of 120 cm$^{-1}$, comfortably greater than a typical mode-width for such a system (40 cm$^{-1}$), see for example~\cite{Long_ACSPhot_2014_2_130,Menghrajani_ACSPhot_2020_7_2448}. 

\subsection{Attenuation: R6G dye in ethanol}

\begin{figure}
\includegraphics[width=1.0\columnwidth]{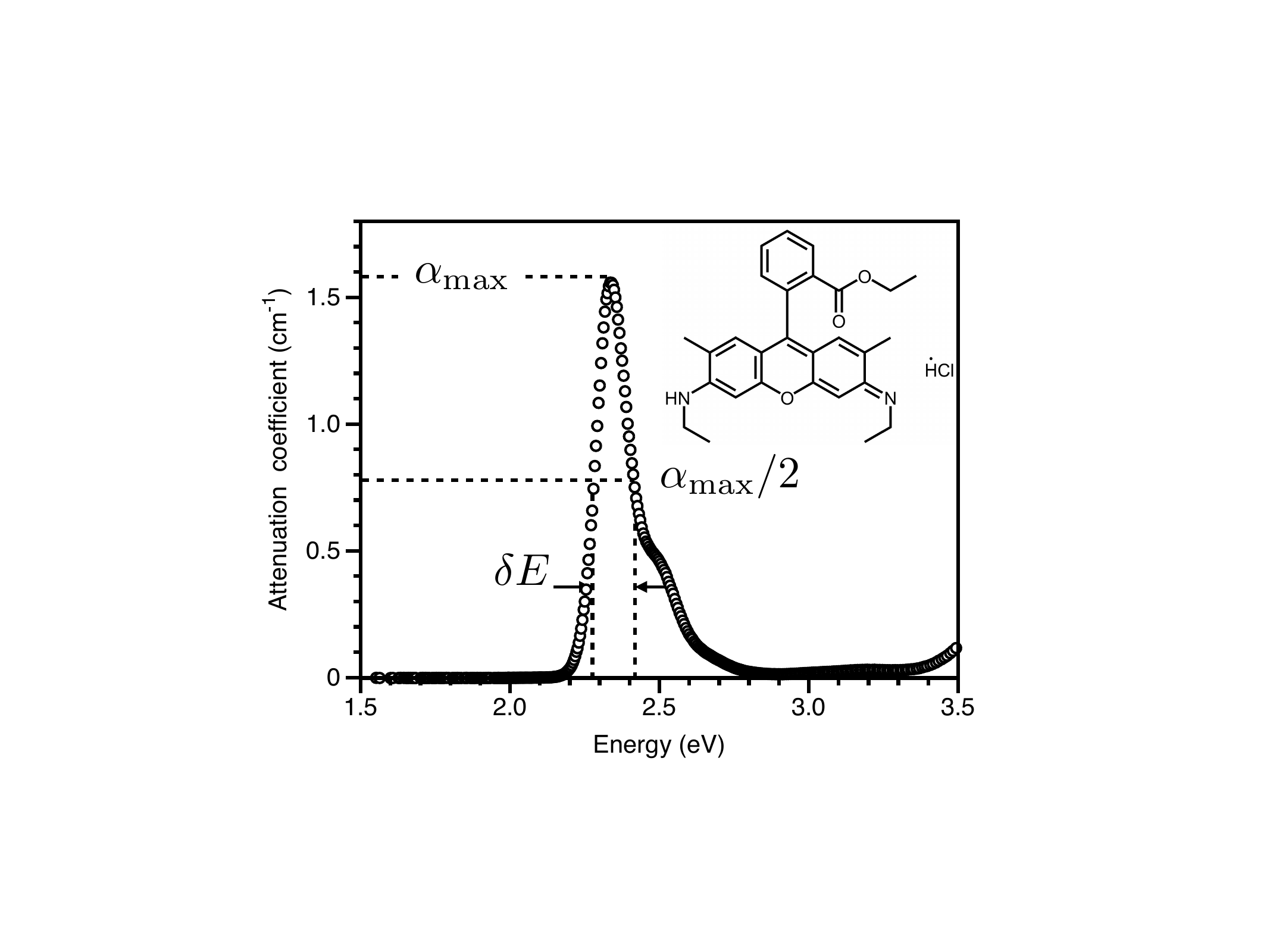}
\caption{\textbf{Attenuation of R6G in ethanol}. The sample is a solution of R6G in ethanol at 6.89 $\mu$M, the sample was held in a cuvette of 1 cm path length. The inset shows the chemical structure of R6G.}
\label{fig:R6G atten}
\end{figure}

Our final example is that of a dye in solution, in this case R6G in ethanol at 6.89 $\mu$M. The measured attenuation coefficient is shown in FIG.~\ref{fig:R6G atten}.
The data are in terms of eV, see FIG.~\ref{fig:TDBC}, the formula we need for the coupling strength is in row two, right hand column of Table~\ref{tab:params},
\begin{equation}
g_N = \frac{1.78\times 10^{-3}}{\sqrt{n_{\textrm{host}}}}\sqrt{\alpha_{max}E\,\delta E}.
\end{equation}
%
From FIG.~\ref{fig:R6G atten} the maximum attenuation is 1.6 cm$^{-1}$, the width is $\delta E \sim$ 0.14 eV, and the transition frequency is 2.34 eV. Again assuming $\sqrt{n_{\textrm{host}}}\,\sim 1.2$, we find $g_N\,\sim$ 1 $\times 10^{-3}$ eV, clearly orders of magnitude below what is required for strong coupling. However, these data are for a very dilute solution, much higher concentrations are used in thin films. As an example, Hakala \textit{et al.}~\cite{Hakala_PRL_2009_103_053602}
used a concentration of 200 mM. If we assume no photo-physical properties are altered as the concentration increases then this gives a value for the interaction strength of $g_N=$ 0.12 eV. With a transition linewidth of $\Gamma\,\sim\,\delta E\,\sim$ 0.14 eV, the strong coupling condition indicates that provided we can use a cavity mode with a width $K\,<$ 0.34 eV it should be possible to observe strong coupling. The surface plasmon mode used by Hakala \textit{et al.} had an estimated width of $<$ 0.05 eV, easily satisfying the required criterion.\\

\section{Conclusions and Discussion}

We have presented an analysis of the strong coupling interaction between a confined light field and an ensemble of molecules that links the coupling strength to readily measured parameters. This analysis should enable, for example, a UV-VIS spectrometer to be sufficient to determine whether a material might in principle show strong coupling. This framework should thus enable the development and evaluation of new candidate molecular materials for strong coupling. Before closing we should note some restrictions of our approach. First, we have assumed that the molecules of interest fill the mode volume - as we noted above, this is often not the case, in which situation our predictions will overestimate the extent of any strong coupling. Second, we have assumed that the vacuum electric field strength is constant across all molecules, for small (e.g. plasmonic) structures this may not be the case. Third, we have assumed we can consider a bulk material response to be appropriate. For situations where only a few molecules, perhaps only one molecule, are involved, an alternative approach will be required. Fourth, we have assumed that the resonance of interest is spectrally well isolated from other (molecular) resonances. Again, this may not be the case, especially for example in light-emitting organic materials. Fifth, we have ignored any complications due to variations in alignment between the electric field of the cavity mode and the orientation of the molecular dipole moments. Despite these restrictions we think the approach we have presented here provides a useful way to compare different candidate materials, and to estimate best-case scenarios re: the extent of strong coupling.\\

\section*{Acknowledgements}
We are especially grateful to Prof. Neil Hunter, Dr. Jenny Clark, Prof. Nick Williams and Dr. Kishan Menghrajani for many useful discussions. \\

\section*{Research funding}
All authors acknowledge the financial support of the UK EPSRC through grant EP/T012455/1, `Molecular Photonic Breadboards' (www.breadboards.org). WLB acknowledges the support of European Research Council through the `photmat' project (ERC-2016-AdG-742222:www.photmat.eu).

\section*{Author contributions}
MSR, GJL and WLB conceived the project. MSR and WLB undertook the theoretical analysis, and wrote the manuscript with input from all authors. ECJ, WPW, DB, RHG and RDJO made samples and undertook spectroscopy to provide data for the worked examples, with input from SPA. All authors have accepted responsibility for the entire content of this manuscript and approved its submission.



\bibliography{bulk_new_refs}

\appendix 
%

\section{Strong coupling criteria}
\label{app:criteria}

There are various forms for the strong coupling condition. One well-used condition is~\cite{Rider_CP_2021_62_217},
\begin{align}
g^2_N > \frac{\gamma^2_\mathrm{mol}+\gamma^2_\mathrm{cav}}{2},
\label{eq:sc condition 1}
\end{align}
where $\gamma_\mathrm{mol}$ and $\gamma_\mathrm{cav}$ are the dephasing rates of the molecular transition and cavity field respectively (see comment in Appendix C). A more convenient condition from an experimental point of view can be given in terms of the linewidths rather than the dephasing rates. If we note that the extent of the anti-crossing is $\Omega_R=2g_N$ (this is also known as the Rabi frequency), and that the molecular and cavity linewidths, $\Gamma$ and $K$ respectively, are typically twice the dephasing rates~\cite{Rider_CP_2021_62_217,Barnes_JOpt_2024} then we can write the strong coupling condition as~\cite{Rider_CP_2021_62_217} ,
\begin{align}
\Omega_R=2g_N > \frac{\Gamma+K}{2}.
\label{eq:sc condition 3}
\end{align}
This condition is used in the present work.

\section{Interaction strength in terms of dipole moment}
\label{app:dipole}

We can make use of equation \eqref{eq:interaction_2} for the interaction strength in terms of the dipole moment to provide the following worked example. We choose as our system the coupling of the C=O bind vibration in the polymer PVA to a planar infra-red optical microcavity mode, following the pioneering work of Shalabney \textit{et al.} \cite{Shalabney_NatComm_2015_6_5981}, and Long and Simpkins~\cite{Long_ACSPhot_2014_2_130}. Elsewhere~\cite{Barnes_JOpt_2024} we use a transfer matrix approach to model the transmission of the bare PVA film used by Shalabney \textit{et al.,} \cite{Shalabney_NatComm_2015_6_5981}. Doing so we are able to determine the dipole moment to be 0.97 $\times 10^{-30}$ Cm. For PVA we have that $N/V$ = 8.33 $\times$ 10$^{27}$ m$^{-3}$, and $\varepsilon_{\textrm{host}}$ = 2. For the C=O resonance $\omega_0$ = 3.26 $\times$ 10$^{14}$ rad s$^{-3}$~\cite{Barnes_JOpt_2024}. Putting these values into \eqref{eq:interaction_1} we find $g_N$ = 80 cm$^{-3}$. Note that to find this value we divided the right-hand side of \eqref{eq:interaction_1} by $\sqrt{3}$ to take account of the random orientation of dipole moments expected in this system, see section Appendix~\ref{app:mat}. This compares with the measured value~\cite{Shalabney_NatComm_2015_6_5981} of $g_N$ $\approx$ 85 cm$^{-3}$.

\section{Interaction strength and material response}
\label{app:mat}
\begin{equation}
\label{eq:LO_1}
\varepsilon(\omega) = \varepsilon_{\textrm{host}}+\frac{f_p\,\omega_p^2}{\omega_0^2-\omega^2-i\omega\gamma},
\end{equation}
where $f_p$ is the oscillator strength, $\varepsilon_{\textrm{host}}$ is the background permittivity and, as noted above, $\gamma$ is the dephasing (loss) rate. The oscillator strength $f_p$ is related to the dipole moment and is given by~\cite{NandH},
\begin{equation}
f_p = \frac{2\, m_e\, \omega_0}{3\, \hbar\, e^2}|\mu|^2,
\label{eq:osc strength plas freq}
\end{equation}
%
Note that equation \eqref{eq:osc strength plas freq} is based on the assumption that the dipole moments are randomly oriented in space, if they aligned, i.e. they are all parallel to each other, then the factor of 3 in the denominator should be removed. We can re-write equation \eqref{eq:osc strength plas freq} to find the dipole moment as,
\begin{equation}
|\mu| = \sqrt{\frac{3\, \hbar\, e^2}{2\, m_e\, \omega_0}f_p}.
\label{eq:dip_mom_vac}
\end{equation}
The parameter $\omega_p$ in \eqref{eq:LO_1} is given by~\cite{BandH},
\begin{equation}
\omega_p^2 = \frac{N e^2}{V\varepsilon_0\, m_e},
\label{eq:plas freq}
\end{equation}
where $N/V$ is the density of molecules, $e$ is the electronic charge, and $m_e$ is the electron mass. (If the electrons were free this $\omega_p$ would be the plasma frequency. In the materials we consider here the electrons are not free, but the nomenclature has stuck, we can simply treat $\omega_p$ as a short cut for the parameters given in \eqref{eq:plas freq}.) Note that $\omega_p$ contains the density of the molecules, it is a material-specific property. Substituting \eqref{eq:dip_mom_vac} and \eqref{eq:plas freq} into \eqref{eq:interaction_1} we find,
\begin{equation}
\label{eq:gn plas freq_aligned}
g_{N,{\textrm{ aligned}}} = \frac{\sqrt{3}}{2}\sqrt{\frac{f_p\, \omega_p^2}{\varepsilon_{\textrm{host}}}},
\end{equation}
However, recall that equation \eqref{eq:interaction_1} is based on the dipole moments being aligned with the cavity field. If instead the dipole moments are randomly oriented then the factor of $\sqrt{3}$ in \eqref{eq:gn plas freq_aligned} needs to be removed, so that,\\
\begin{equation}
\label{eq:gn plas freq_random}
g_{N,{\textrm{ random}}} = \frac{1}{2}\sqrt{\frac{f_p\, \omega_p^2}{\varepsilon_{\textrm{host}}}},
\end{equation}
From here on when we write $g_N$ we will imply $g_{N,{\textrm{ random}}}$ unless indicated otherwise. \\

In Appendix~\ref{app:ext} we transform \eqref{eq:gn plas freq_random} into a version based on more easily measurable parameters. First let us note that the numerator in equation \eqref{eq:LO_1} is often replaced with $f_0\,\omega_0^2$~\footnote{This is usually done for convenience, the molecular resonance frequency $\omega_0$ is accessible through experiment, whilst $\omega_p$ is not.}. In this case \eqref{eq:gn plas freq_random} is replaced with,
\begin{equation}
\label{eq:gn mol freq}
g_N = \frac{1}{2}\sqrt{\frac{f_0\, \omega_0^2}{\varepsilon_{\textrm{host}}}},
\end{equation}
where now $f_0$ is given by,
\begin{align}
\label{eq:osc strength mol feq}
   f_0 &= \frac{2 N}{3\,V\,\varepsilon_0 \hbar\omega_0}|\mu|^2, 
\end{align}
and where again the molecular dipole moments are assumed to be randomly oriented.

\section{Derivation of the interaction strength in terms of the extinction coefficient}
\label{app:ext}

On resonance $\omega=\omega_0$ and the coefficients $\kappa$, $\alpha$ etc. take their maximum values. We can then write equation \eqref{eq:LO_1}, the permittivity using the Lorentz oscillator, as,
\begin{align}
\label{eq:varepsilon_res}
\varepsilon(\omega_0) = \varepsilon_{\textrm{host}} + i\frac{f_p\omega^2_p}{\gamma\omega_0},
\end{align}
and, making use of equation \eqref{eq:gn plas freq_random}, we can write the permittivity as,
\begin{align}
\label{eq:varepsilon_res_gN}
\varepsilon(\omega_0) = \varepsilon_{\textrm{host}} + i\frac{4\,\varepsilon_{\textrm{host}} g_N^2}{\gamma\omega_0},
\end{align}
The extinction coefficient is the imaginary component of the refractive index, $\kappa$ and,
since the permittivity and refractive index are related by $\varepsilon(\omega)=(n+i\kappa)^2$, we can find an expression for $\kappa$ in terms of $\varepsilon(\omega)$ by finding the imaginary part of the square root of equation \eqref{eq:varepsilon_res_gN}, i.e.
\begin{align}
\label{eq:extinction_as_root_1}
\kappa_{\textrm{max}} &= \sqrt{\frac{|\varepsilon_r(\omega_0)|-\mathrm{Re}[\varepsilon_r(\omega_0)]}{2}}.
\end{align}
where $\varepsilon_r(\omega_0)$ is the real part of $\varepsilon(\omega_0)$. Using \eqref{eq:varepsilon_res_gN} and \eqref{eq:extinction_as_root_1} we can write the extinction coefficient as,
\begin{align}
\label{eq:extinction_as_root_2}
    \sqrt{2}\kappa_{\textrm{max}}= \sqrt{\varepsilon_{\textrm{host}}\left|1+\left(\frac{4\,g_N^2}{\gamma\,\omega_0}\right)^2\right|-\varepsilon_{\textrm{host}}}.
\end{align}
We chose to avoid the ultrastrong coupling regime~\cite{Kockum_NatRevPhys_2019_1_19}, for which $g_N\geq\omega_0/10$, so that $g_N<\omega_0$ and $g_N\leq\gamma$. This allows us to use the binomial theorem to write $|\varepsilon(\omega_0)|=[1 + 1/2(4\,g_N^2/\omega_0\gamma)^2]$, so that now \eqref{eq:extinction_as_root_2} can be re-written to give the extinction coefficient $\kappa$ as,
\begin{align}
\label{eq:extinction_as_root_3}
\sqrt{2}\kappa_{\textrm{max}}= \sqrt{\frac{\varepsilon_{\textrm{host}}}{2}\left(\frac{4g_N^2}{\gamma\,\omega_0}\right)^2}.
\end{align}
Rearranging we can write the interaction strength in terms of the extinction coefficient as,
\begin{align}
\label{eq:gn extinction 2}
g_N=\frac{1}{2\sqrt{n_{\textrm{host}}}}\sqrt{\kappa_{\textrm{max}}\,\omega_0\,\delta\omega},
\end{align}
in which $\delta\omega\sim 2\gamma$~\cite{Barnes_JOpt_2024}. Note that we have assumed an isotropic distribution of dipole moments with respect to the (cavity mode) electric field. If the dipole moments are aligned with the field, then because $g_N\propto f_P$ we will need to multiply $g_N$ in equation \eqref{eq:gn extinction 2} by $\sqrt{3}$. We show below that a very similar expression can be derived through a more formal spectroscopic approach that does not rely on the assumption that the molecular resonance is well-described by the Lorentz Oscillator model, there is simply a small change in the prefactor to equation \eqref{eq:gn extinction 2}, it becomes,
\begin{align}
\label{eq:gn extinction 2 b}
g_N=\frac{0.56}{\sqrt{n_{\textrm{host}}}}\sqrt{\kappa_{\textrm{max}}\,\omega_0\,\delta\omega}.
\end{align}
It is this version that we use here, and that is reproduced in Table~\ref{tab:params}.

\section{Different forms of the Beer-Lambert law}
\label{app:Beer-Lambert}

\begin{figure}[h!]
\centering
\includegraphics[width=0.9\columnwidth]{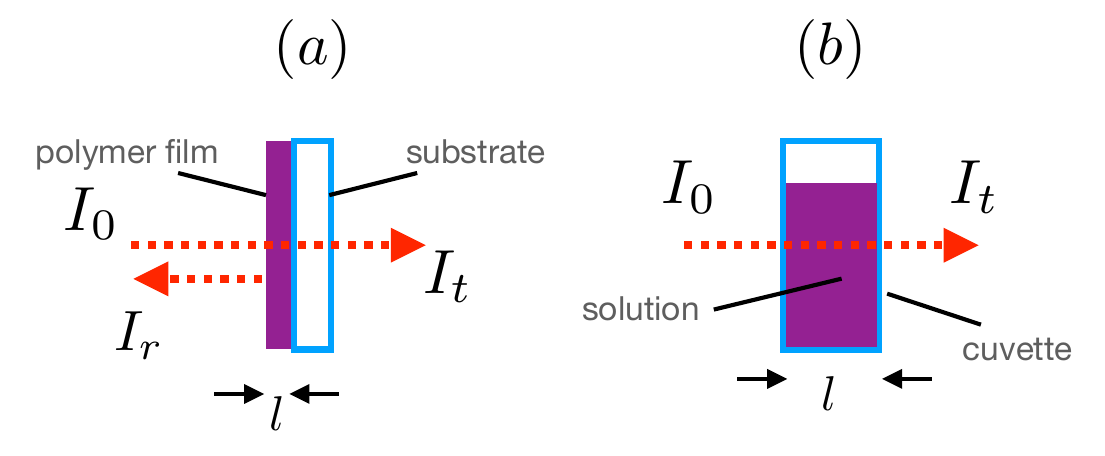}
\caption{\textbf{Typical Beer-Lambert set-up} to measure the extinction of a solution/solid thin film sample. On the left (a) a thin dye-doped polymer film is shown, on the right (b) a cuvette containing a dye solution. The intensity of light transmitted by a sample of path length (thickness) $l$ is measured as a function of the frequency/wavelength of the light. Note that a number of precautions need to be taken to ensure transmittance data are suitable for this kind of analysis~\cite{MolFluo}. In the case of thin films of strongly absorbing molecules, account needs to be taken of the extent of any reflected power produced by the sample, see main text.}
\label{fig:Beer-Lambert}
\end{figure}

Frequently the materials of interest -- e.g. dye-doped polymers or solutions -- are initially characterised by measuring a transmission spectrum and from the data so acquired one of the following parameters determined: the extinction coefficient, $\kappa(\omega)$; the attenuation coefficient, $\alpha(\omega)$; the molar absorption coefficient, $\epsilon(\omega)$; and the optical density, $\textrm{OD}$. These parameters are extracted from experimental data using various (equivalent) forms of the Beer-Lambert law~\cite{MolFluo} typically using a set-up similar to that shown in FIG.~\ref{fig:Beer-Lambert}. Here we look at each version of the Beer-Lambert in turn.
We begin by looking at the attenuation coefficient, $\alpha(\omega)$. 

\subsection{Attenuation coefficient, $\alpha$}

In terms of the attenuation coefficient, $\alpha$, the Beer-Lambert law takes the form,
\begin{equation}
\label{eq:atten_coeff_1}
\textrm{I}(z) = \textrm{I}(0)\,e^{-\alpha z}, 
\end{equation}
so that,
\begin{equation}
\label{eq:atten_coeff_2}
\alpha = \frac{1}{t}\,\textrm{log}_e\,\left({\frac{\textrm{I}(0)}{\textrm{I}(t)}}\right), 
\end{equation}
so that, combining \eqref{eq:extinc_coeff_4} and \eqref{eq:atten_coeff_2}, we have,
\begin{equation}
\label{eq:atten_extinc}
\alpha=2k_0\kappa.
\end{equation}

\noindent Note that we take for the units of $\alpha$ cm$^{-1}$.

\subsection{Molar absorption coefficient, $\epsilon$}

A common measurement parameter is the molar absorption coefficient~\footnote{It is often referred to as the molar extinction coefficient, but this is not the approved term~\cite{Braslavsky_PAC_2007_79_293}.} $\epsilon$, and is given by~\cite{Turro,Atkins_and_Freidman_4},
\begin{equation}
\label{eq:mol_ext_coeff}
\epsilon = \rm{log}_{10}\left(\frac{I(0)}{I(\rm{t})}\right)\frac{1}{\textit{l}\,C_m}.
\end{equation}
In calculating numerical quantities using the correct units is vital, and this applies to all of the different quantities discussed above. We will look at each of them in the notes below, but a word about the molar absorption coefficient here will be helpful. The units for the path length, $l$, are cm, whilst the units for the molecular concentration, $C_m$, are moles dm$^{-3}$, i.e. moles per litre, whilst the molar absorption coefficient usually has units of dm$^{3}$ mol$^{-1}$ cm$^{-1}$.\\

\subsection{Extinction coefficient, $\kappa$}

With reference to FIG.\ref{fig:Beer-Lambert}, the incident, $\textrm{E(0)}$, and transmitted, $\textrm{E}(z)$, electric fields are related by,

\begin{equation}
\label{eq:extinc_coeff_1}
   \textrm{E}(z) = \textrm{E}(0)\,e^{-ikz}, 
\end{equation}

\noindent where the wavevector, $k$ and the complex index of refraction, $(n+i\kappa)$, of the material being measured are related through $k=k_0(n+i\kappa)$.
It is the $\kappa$ term in $(n'+i\kappa)$ that is responsible for the attenuation of the light, so we then have,
\begin{equation}
\label{eq:extinc_coeff_2}
   \textrm{E}(z) = \textrm{E}(0)\,e^{-k_0 \kappa z}, 
\end{equation}
so that in terms of intensity,
\begin{equation}
\label{eq:extinc_coeff_3}
   \textrm{I}(z) = \textrm{I}(0)\,e^{-2k_0 \kappa z}, 
\end{equation}
giving,
\begin{equation}
\label{eq:extinc_coeff_4}
   \kappa = \frac{1}{2\,k_0\,t}\,\textrm{log}_e\,\left({\frac{\textrm{I}(0)}{\textrm{I}(t)}}\right), 
\end{equation}
where $t$ is the sample thickness in metres, and in the interests of clarity we have specified the base for the logarithm. Regarding units, $\kappa$ is dimensionless, so there are no units to worry about.

\subsection{Absorbance, $a$}

The absorbance (also known as optical density, OD) is a variant of the attenuation coefficient. The the absorbance \textit{a} is defined as,
\begin{equation}
\label{eq:OD_1}
\textit{a} = \textrm{OD}= \textrm{log}_{10}\,\left({\frac{\textrm{I}(0)}{\textrm{I}(t)}}\right).
\end{equation}
The optical density (absorbance) and the attenuation coefficient are thus related by,
\begin{equation}
\label{eq:OD_2}
\textit{a}=\textrm{OD}= \frac{\alpha\,t}{\,\textrm{log}_{e}(10)}.
\end{equation}

Regarding units, $a$ is dimensionless, so there are no units to worry about.

\subsection{Interaction strength in terms of optical parameters.}

The equations in Table~\ref{tab:params} are based on the different variants of the Beer-Lambert law given above.\\

To connect these different parameters ($\epsilon,\, k\,\alpha,\,a$) with the interaction strength we need to express them in terms of the oscillator strength. A useful starting point is to relate the molar absorption coefficient and the oscillator strength. Kuhn \textit{et al.}~\cite{Kuhn_Forsterling_Waldeck} and Valeur and Berberan-Santos~\cite{MolFluo} provide details, the result is,
\begin{equation}
\label{eq:f_and_mol_ext_coeff_1}
\frac{f_p}{n_{\textrm{host}}} = \frac{4\,\textrm{log}_{10}\,\varepsilon_0\,c\,m_e}{N_A\,e^2}\int\epsilon(\nu)\,d\nu,
\end{equation}
where $\nu$ is the frequency in Hz. Note that the factor of $n_{\textrm{host}}$ in the denominator on the l.h.s of this equation is there to take account of the way the energy density inside the molecular material is different from that in free space, see equation 9.29 in~\cite{Kuhn_Forsterling_Waldeck}. Using standard values for the fundamental constants, the prefactor in equation \eqref{eq:f_and_mol_ext_coeff_1} can be evaluated so that and is found to be, $f_p = 1.47 \times 10^{-18}\int\epsilon(\nu)\,d\nu$. Because the units for $\epsilon$ are dm$^{3}$ mol$^{-1}$ cm$^{-1}$, we need to divide by a factor of 10 to obtain,\footnote{For an alternative derivation based on the Einstein coefficients, see~\cite{Methods_3}.},
\begin{equation}
\label{eq:f_and_mol_ext_coeff_2}
\frac{f_p}{n_{\textrm{host}}} = 1.44 \times 10^{-19}\int\epsilon(\nu)\,d\nu.
\end{equation}
In photochemistry it is common to use a range of parameters for the spectral measurement, Hz is just one choice. Other choices include wavenumber, $\bar{\nu}$, (cm$^{-1}$), angular frequency, $\omega$, (rad s$^{-1}$) and energy, (eV). In terms of these other units, the oscillator strength is,
\begin{equation}
\begin{aligned}
\label{eq:f_and_mol_ext_coeff_var}
f_p & = 4.32 \times 10^{-9}\,n_{\textrm{host}}\,\int\epsilon(\bar{\nu})\,d\bar{\nu} \\
& = 2.29 \times 10^{-20}\,n_{\textrm{host}}\,\int\epsilon(\omega)\,d\omega \\
& = 3.48 \times 10^{-5}\,n_{\textrm{host}}\,\int\epsilon(\omega)\,d(eV).
\end{aligned}
\end{equation}
Equation~\eqref{eq:f_and_mol_ext_coeff_var} can often be approximated as (see~\cite{Turro} equation 5.40),
\begin{equation}
\label{eq:mol_ext_coeff_2}
\begin{aligned}
f_P & \approx 4.32 \times 10^{-9}\,n_{\textrm{host}}\,\,\epsilon_{\rm{max}}\, \delta\bar{\nu} \\
& \approx 2.29 \times 10^{-20}\,n_{\textrm{host}}\,\,\epsilon_{\rm{max}}\,\delta\omega \\
& \approx 3.48 \times 10^{-5}\,n_{\textrm{host}}\,\,\epsilon_{\rm{max}}\,\delta(eV),
\end{aligned}
\end{equation}
where $\delta\bar{\nu}$, $\delta\omega$, $\delta(eV)$ are the spectral widths (FWHM) of the extinction feature. Noting that the number density of molecules (number per cubic metre) $N/V=10^3\,C_m\,N_A$, where $N_A$ is Avagadro's number, then we can use \eqref{eq:mol_ext_coeff_2} and \eqref{eq:plas freq} in \eqref{eq:gn plas freq_random}, together with known values of physical constants, to find $e^2/{\varepsilon_0\,m_e}=3.18 \times 10^3$, so that,
\begin{equation}
\label{eq:gn mol_ext_coeff}
\begin{aligned}
g_N & = \frac{0.24}{\sqrt{n_{\textrm{host}}}}\sqrt{\epsilon_{max}\,C_m\, \delta\bar{\nu}} \\
& = \frac{1.05 \times 10^{5}}{\sqrt{n_{\textrm{host}}}}\sqrt{\epsilon_{max}\,C_m\, \delta\omega} \\
& = \frac{2.70 \times 10^{-3}}{\sqrt{n_{\textrm{host}}}}\sqrt{\epsilon_{max}\,C_m\, \delta(eV)},
\end{aligned}
\end{equation}
where, for clarity, the maximum (spectral peak) value of the molar absorption coefficient, $\epsilon_{max}$ is found from transmittance data using equation \eqref{eq:mol_ext_coeff}, the concentration $C_m$ is in moles per litre, and the spectral width of the extinction peak, $\delta\bar{\nu}, \delta\omega$, $\delta(eV)$ are in wavenumbers, metres, radians per second, and eV (as appropriate), and $n_{\textrm{host}}$ is the background refractive index, e.g. of the solvent.\\


Now that we have what we need for the molar absorption coefficient, we can use equations~\eqref{eq:gn extinction 2 b},~\eqref{eq:atten_extinc} and~\eqref{eq:OD_2} to find similar expressions in terms of the maximum extinction coefficient, $\kappa_{max}$, the maximum absorption coefficient, $\alpha_{max}$, and the maximum absorbance, \textit{a}$_{max}$, which are related according to,
\begin{equation}
\label{eq:various_coefficients_1}
\epsilon_{max}\,C_m = \frac{\alpha_{max}}{2.3} = \frac{2\kappa_{max}\,\omega_0}{2.3\,\times\,10^2\,c} = \frac{\textit{a}_{max}\,\textrm{log}_e(10)}{2.3\,l}.
\end{equation}

\noindent where the attenuation coefficient, $\alpha$, is in units of cm$^{-1}$, and the path length, $l$, is also in cm.

\section{Details of the synthesis and characterisation of Nile Red analogue, PAGEO5MA brushes, and the Nile Red-PAGEO5MA brush system}
\label{app:Methods}
\subsection{Materials}
Sodium periodate ($\geq$99.8\%, NaIO$_4$), (3-aminopropyl)triethoxysilane ($>$99\%, APTES), triethylamine (99\%, NEt$_3$), 2-bromoisobutyryl bromide ($>$99\%, BiBB), copper(II) chloride (99.999\%, CuCl$_2$), ascorbic acid ($>$98\%, AscA), sodium cyanoborohydride (95\%, NaBH$_3$CN), dichloromethane ($>$99\%, DCM), diethyl ether ($>$=99.8\%), \emph{N}-(1-naphthyl)ethylenediamindihydrochloride ($>$98\%) and ammonia solution (NH$_4$OH, 35\%) were purchased from Sigma-Aldrich, UK and were used without further purification. 3-Diethylaminophenol (99\%, Acros Organics, UK), sodium azide ($>$99\%, Acros Organics, UK), sodium nitrite ($>$97\%, Alfa Aesar, UK), dibromoethane (98\%, Alfa Aesar, UK), hydrochloric acid (HCl, 35 wt\%, Scientific Laboratory Supplies, UK), potassium carbonate (100\%, K$_2$CO$_3$, VWR, UK), petroleum ether (40-60, 95\% VWR), magnesium sulfate (dried, Fisher Scientific, UK), sodium hydroxide ($>$97\%, NaOH, Fisher Scientific, UK), sodium chloride ($>$=99.5\%, NaCl, Fisher Scientific, UK) were also used as purchased. GEO5MA monomer was kindly donated by GEO Specialty Chemicals, UK and was used without further purification. All other solvents were purchased from Fisher Scientific, UK and were used as received unless otherwise stated herein. N,N,N',N'',N''-Pentamethyldiethylenetriamine ($>$99\%, PMDETA) was also purchased from Fisher Scientific, UK. Deionized water (pH 6.8) was obtained using an Elga Elgastat Oprion 3A water purification system. Native oxide-coated silicon wafers were purchased from Pi-KEM, UK. Column chromatography was performed using silica gel (40-60 $\mu$m, VWR, UK) as the stationary phase, monitored by thin layer chromatography (TLC) using Merck silica gel 60 F254 plates and visualised under UV light if required. NMR spectra were recorded on a Bruker Avance III HD and chemical shifts referenced to residual solvent. Mass spectra were recorded by the University of Sheffield Mass Spectrometry facility on an Agilent 6530 Accurate Mass LC-MS QToF mass spectrometer (LCMS, high resolution) or a Waters LCT Classic (direct infusion, low resolution). 3-Diethylamino-2-nitrosophenol hydrochloride~\cite{Brousmiche2004} and 2-Hydroxy Nile Red~\cite{Ghini2009} were synthesised according to published procedures. Microscope coverslip glass slides (22 mm, 50 mm, 1.5 mm thickness) were obtained from Menzel-Gläser, Germany. Ethyl acetate (EtOAc, 99\%) and dichloromethane (DCM,99\%) were purchased from Fisher Science and used as provided.

\subsection{Preparation of 2-(2-aminoethoxy)Nile Red}

\begin{figure*}
\includegraphics[width=\textwidth]{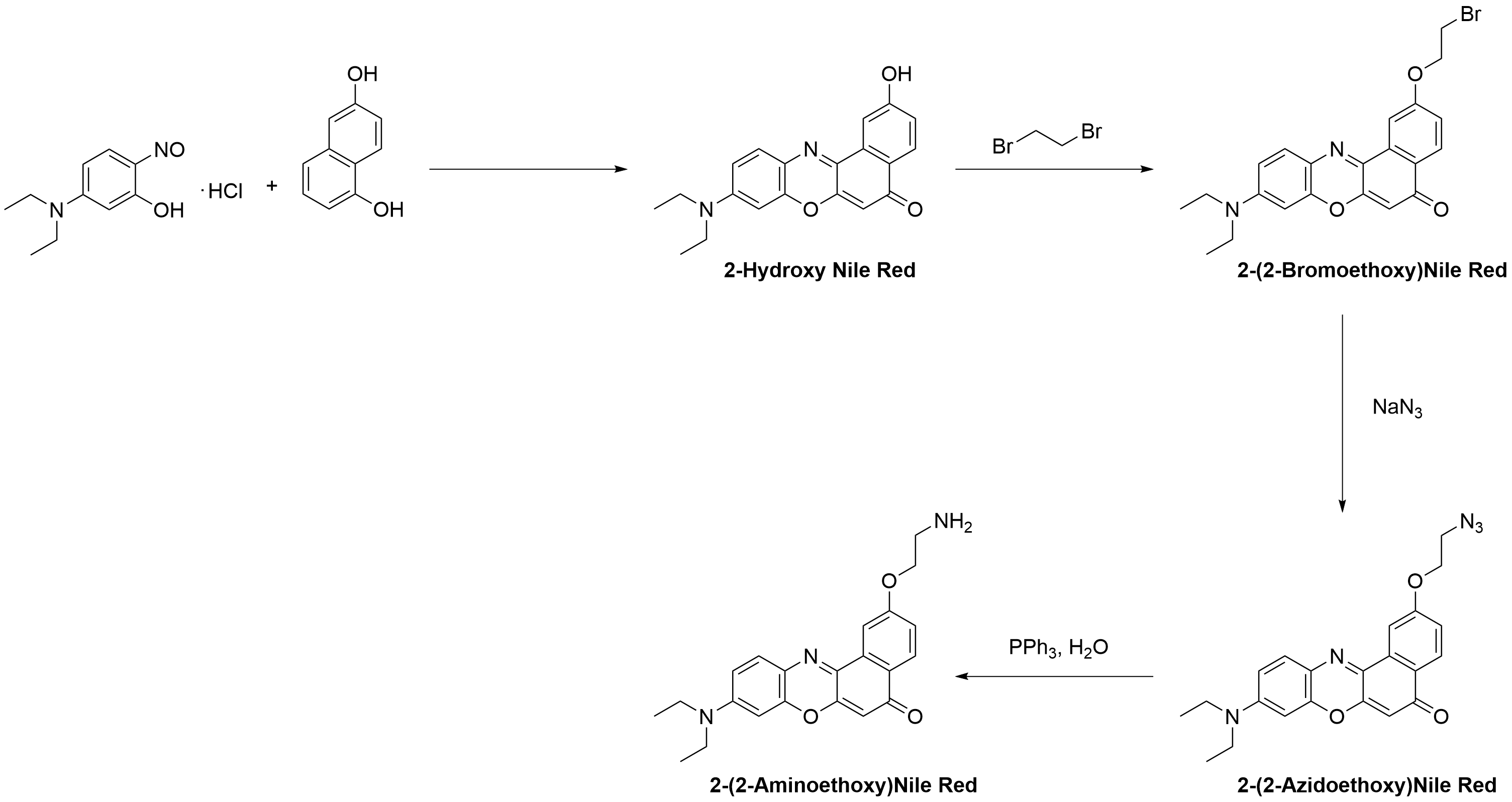}
\caption{\textbf{Synthesis of 2-(2-aminoethoxy)Nile Red} }
\label{fig:NR_Synth}
\end{figure*}

\subsubsection{Preparation of 2-(2-bromoethoxy)Nile Red}
Our synthesis route is a version of a previously reported protocol~\cite{Brousmiche2004}. 2-Hydroxy Nile Red (0.407 g, 1.22 mmol), potassium carbonate (1.686 g, 12.20 mmol) and anhydrous DMF (10 mL) were combined under N$_2$. 1,2-Dibromoethane (3.2 mL, 37.13 mmol) was added to the stirred mixture and the reaction was heated to 65 $^\circ$C. The reaction was tracked by TLC (1:1 EtOAc/pet ether) and cooled to 20 $^\circ$C after 1 h 10 min. DCM (50 mL) and Deionized water (50 mL) were added, the layers separated and the organic layer washed with DI water (2 $\times$ 30 mL), the aqueous washings back-extracted with DCM (3 $\times$ 30 mL) and the combined organic layers washed with brine (80 mL), dried over MgSO$_4$ and concentrated in vacuo. The crude solid was subjected to flash chromatography (1:1 EtOAc/pet ether, rising to 2:1 once fast moving impurities had eluted) yielding 2-(2-bromoethoxy)Nile Red as a purple solid (0.292 g, 0.66 mmol, 54\%): $^1$H NMR (400 MHz, CDCl$_3$) $\delta$ 8.27 (d, J = 8.7 Hz, 1H), 8.10 (d, J = 2.6 Hz, 1H), 7.64 (d, J = 9.1 Hz, 1H), 7.23 (dd, J = 8.7, 2.6 Hz, 1H), 6.70 (dd, J = 9.1, 2.8 Hz, 1H), 6.50 (d, J = 2.7 Hz, 1H), 6.34 (s, 1H), 4.54 (t, J = 6.2 Hz, 2H), 3.76 (t, J = 6.2 Hz, 2H), 3.50 (q, J = 7.1 Hz, 4H), 1.29 (t, J = 7.1 Hz, 6H); $^{13}$C\{$^1$H\} NMR (100 MHz, CDCl$_3$) $\delta$ 183.09, 160.67, 152.09, 150.82, 146.87, 139.67, 134.08, 131.10, 127.91, 126.17, 124.71, 118.32, 109.61, 106.65, 105.28, 96.28, 77.37, 77.26, 77.05, 76.73, 68.06, 45.11, 28.97, 12.65; HRMS (ESI): m/z 441.0818 [M+H]$^+$, calculated for C$_{22}$H$_{22}$N$_2$O$_3$Br$^+$ 441.0808.

\subsubsection{Preparation of 2-(2-azidoethoxy)Nile Red}
Our synthesis route is a version of a previously reported protocol~\cite{Ghini2009}. 2-(2-Bromoethoxy)Nile Red (0.290 g, 0.657 mmol) and NaN$_3$ (0.064 g, 0.984 mmol) were dissolved in anhydrous DMF (10 mL). The mixture was stirred at 80 $^\circ$C under N$_2$ for 20 h and cooled to RT. EtOAc (50 mL) was added and the solution washed with 3M NaOH (3 $\times$ 50 mL) and brine (50 mL). The organics were dried over MgSO$_{4}$ and concentrated in vacuo, yielding 2-(2-azidoethoxy)Nile Red as a purple solid (0.261 g, 0.647 mmol, 98\%). This was used without further purification: $^1$H NMR (400 MHz, CDCl$_3$) $\delta$ = 8.24 (d, J = 8.7 Hz, 1H), 8.05 (d, J = 2.7 Hz, 1H), 7.58 (d, J = 9.0 Hz, 1H), 7.20 (dd, J = 8.7, 2.6 Hz, 1H), 6.66 (dd, J = 9.1, 2.7 Hz, 1H), 6.45 (d, J = 2.7 Hz, 1H), 6.30 (s, 1H), 4.38 (t, J = 4.9 Hz, 2H), 3.71 (t, J = 5.0 Hz, 2H), 3.48 (q, J = 7.1 Hz, 4H), 1.28 (t, J = 7.1 Hz, 6H) ppm; $^{13}$C{$^1H$} NMR (100 MHz, CDCl$_3$) $\delta$ = 183.11, 160.80, 152.09, 150.82, 146.87, 139.69, 134.08, 131.09, 127.89, 126.16, 124.71, 118.38, 109.60, 106.44, 105.29, 96.28, 67.24, 50.18, 45.10, 12.64 ppm; HRMS (ESI): m/z 404.1722 [M+H]$^+$, calcd for C$_{22}$H$_{22}$N$_5$O$_3$$^+$ 404.1717.

\subsubsection{Preparation of 2-(2-aminoethoxy)Nile Red}
Our synthesis route is a version of a previously reported protocol~\cite{Tosi2016}. 2-(2-Azidoethoxy)Nile Red (0.030 g, 0.074 mmol) and PPh$_3$ (0.029 g, 0.111 mmol) were dissolved in anhydrous THF (1 mL) and stirred under N$_2$ for 1 h. DI water (5 drops) was added, the mixture stirred for 18 h and the solvent removed in vacuo. The crude material was subjected to flash chromatography (DCM/Methanol (5 \%)/NH$_4$OH(1 \%) rising to 10 \% Methanol once fast moving impurities had eluted) yielding 2-(2-aminoethoxy)Nile Red as a purple solid (0.022 g, 0.058 mmol, 78\%): $^1$H NMR (400 MHz, CDCl$_3$) $\delta$ 8.20 (d, J = 8.7 Hz, 1H), 8.03 (d, J = 2.6 Hz, 1H), 7.55 (d, J = 9.1 Hz, 1H), 7.16 (dd, J = 8.7, 2.6 Hz, 1H), 6.62 (dd, J = 9.1, 2.7 Hz, 1H), 6.41 (d, J = 2.7 Hz, 1H), 6.27 (s, 1H), 4.20 (t, J = 5.2 Hz, 2H), 3.45 (q, J = 7.1 Hz, 4H), 3.18 (t, J = 5.2 Hz, 2H), 1.26 (t, J = 7.1 Hz, 7H), MS (ESI): m/z 378.2 [M+H]$^+$, calculated for C$_{22}$H$_{24}$N$_3$O$_3$$^+$ 378.2.

\subsection{Surface functionalization of glass cover slips and silicon wafers with ARGET ATRP initiator}
Functionalization of planar silicon wafers and glass cover slips with ARGET ATRP initiator moieties was carried out following a experimental protocol reported by Brotherton et al~\cite{Brotherton2023}. Silicon (100) wafers were cut into small pieces (~1 × 1 cm$^2$) before being UV-ozone cleaned for 30 min using a Bioforce Nanosciences ProCleaner. The wafers were then placed in an open petri dish along with a 3 mL glass vial containing $~$100 µL APTES. The petri dish was then placed in a desiccator, which was subsequently sealed and placed under vacuum. Vapour deposition was allowed to proceed for 30 min before the surfaces were removed and placed in a 110 $^\circ$C oven for 30 min. The wafers were then functionalized by immersion in DCM followed by sequential addition of NEt$_3$ (final concentration = 0.2 mM) and BiBB (final concentration = 0.2 mM), allowing the amidation reaction to proceed for 1 h at 22 $^\circ$C. Finally, the initiator-functionalized silicon wafers were rinsed extensively with ethanol and DI water, before drying under a stream of compressed air.

\subsection{Grafting of PGEO5MA brushes via surface-initiated ARGET ATRP from initiator functionalised surfaces}
Following surface functionalization with initiator moieties, surface initiated-activators regenerated via electron transfer, atom transfer radical polymerisation (SI-ARGET ATRP) was employed to grow PGEO5MA homopolymer brushes from each surface. We have recently reported a detailed description of the protocol synthesis~\cite{Brotherton2023}. Briefly, a GEO5MA:CuCl$_2$:PMDETA:ascorbic acid molar ratio of 1000:1:5:5 was used. Deionized water was the solvent with a final monomer concentration of 45\% v/v. Polymerizations were allowed to proceed for 1 h at 22 $^\circ$C in all cases. Each PGEO5MA-functionalized wafer or cover slip was subsequently rinsed thoroughly with ethanol and DI water and then dried using a stream of N$_2$ gas.

\subsection{Selective oxidation of \emph{cis}-diol-functional PGEO5MA brushes for the synthesis of aldehyde-functional PAGEO5MA brushes}
Following our previously reported experimental protocol for selective oxidation of surface-grafted polymer brushes~\cite{Brotherton2023}, PGEO5MA-functionalized planar silicon wafers were immersed in a 3.0 mg mL$^{-1}$ aqueous solution of NaIO$_4$ for 30 min at 22 $^\circ$C, targeting a degree of oxidation of 100\% in all cases. Each aldehyde functional 'PAGEO5MA' functionalized silicon wafer was subsequently rinsed thoroughly with DI water and then dried using a stream of compressed air.

\subsection{Preparation of Nile Red-PAGEO5MA brushes from aldehyde-functional PAGEO5MA brushes via reductive amination}
Oxidized PAGEO5MA brushes were functionalized with 2-(2-aminoethoxy)Nile Red via reductive amination. A 1.0 g dm$^{-3}$ solution of this amine dye was prepared in  methanol before addition of a 1.5 $\times$ molar excess of NaBH$_3$CN. Brushes were immersed in this solution and allowed to react for 24 h at 50 $^\circ$C. The resulting Nile Red-PAGEO5MA brushes were rinsed extensively with methanol, ethanol and deionized water, and then dried under a stream of compressed air.

\subsection{Characterisation of dry polymer brushes}
\subsubsection{Ellipsometry}
Ellipsometry measurements of dry polymer brushes grown from planar silicon wafers were performed using a J. A. Woollam M-2000 V ellipsometer at a fixed angle of incidence of 75$^\circ$ normal to the sample surface in air at 20 °C. A two-layer model consisting of a native oxide layer and Cauchy layer was used to model the data. Data analysis and modelling were performed using Woollam CompleteEase software. Cauchy constants of A$_n$ = 1.459, B$_n$ = 0.006, and C$_n$ = 0 were used. The ellipsometer set-up allowed a relatively large sampling area of approximately 0.5 cm $\times$ 1 cm, which corresponds to around 50\% of the total area of each brush sample. Brushes grown from glass were assumed to be of equal thickness to that grown from silicon wafers present in the same reaction vial. 

\subsubsection{X-ray photoelectron spectroscopy (XPS)}
XPS analysis of dry polymer brushes grafted from planar silicon wafers was performed using a Kratos Axis Supra spectrometer. Step sizes of 0.50 and 0.10 eV were used to record survey and high-resolution C1s, O1s and N1s spectra, respectively. In each case, spectra were recorded from at least two separate areas for each surface-grafted brush. The XPS data were analyzed using Casa XPS software. All binding energies were calibrated with respect to the C1s saturated hydrocarbon peak at 285.0 eV. The degree of functionalization of the brush was determined by comparing the nitrogen/oxygen atomic ratios using the high resolution spectra and was found to be = 82$\pm$ 5\%. 

\subsubsection{UV-visible absorption}
Measurements were performed on Cary50 spectrophotometer (Agilent Technologies, USA) with a baseline transmission through a cuvette prior to measurements. 
Nile Red functionalised PAGEO5MA brushes on glass coverslips were positioned in a plastic cuvette and absorbance at normal incidence measured over a 300 – 800 nm wavelength range.
A background subtraction of a glass coverslips (measured separately) was performed post measurement. 
For the R6G solutions, transmittance was measured over a 200 – 800 nm wavelength range in a quartz cuvette. The concentrations of the dyes was selected to ensure suitable transmittance was measured. 

\end{document}